\documentclass[12pt]{amsart}

\usepackage{amssymb}
\usepackage[leqno]{subeqnarray}

\begin{document}

\title{Soliton stars in the breather limit}
\author{Satyanad Kichenassamy}
\address{Universit\'e de Reims Champagne-Ardenne\\
         Laboratoire de Math\'e\-matiques (CNRS FRE 3111)\\
         Moulin de la Housse, B. P. 1039\\
         F-51687 Reims, France}
\email{satyanad.kichenassamy@univ-reims.fr}

\date{}

\newtheorem{theorem}{Theorem}
\newtheorem{lemma}[theorem]{Lemma}

\theoremstyle{remark}
\newtheorem{remark}{Remark}

\def\tvi{\vrule height 16 pt depth 8 pt width 0pt}

\newcommand{\ep}{\varepsilon}
\newcommand{\pa}{\partial}
\newcommand{\B}{
\raisebox{-1pt}{\hbox{\vrule\vbox to 9pt{\hrule width
9pt\vfill\hrule}\vrule\,}} }
\newcommand{\RR}{{\mathbb R}}
\newcommand{\CC}{{\mathbb C}}
\newcommand{\NN}{{\mathbb N}}
\newcommand{\Ric}{{\mathrm{Ric}}}
\newcommand{\ord}{{\mathrm{ord}}}

\newcommand{\bQ}{\mathbf{Q}}
\renewcommand{\O}{{\mathcal O}}
\newcommand{\scal}[2]{\langle #1\mathop| #2\rangle}
\newcommand{\norm}[1]{\|#1\|}

\def\tr{\mathop{\rm tr}\nolimits}

\vskip 1em

\hfill{\bf Appeared in:} \emph{Classical and Quantum Gravity},
{\bf 25} (2008) 245004.

\vskip 1em
\begin{abstract}
This paper presents an asymptotic reduction of the
Einstein-Klein-Gordon system with real scalar field (``soliton
star problem''). A periodic solution of the reduced system,
similar to the sine-Gordon breather, is obtained by a variational
method. This tallies with numerical computations. As a
consequence, a time-periodic redshift for sources close to the
center of the star is obtained.
\end{abstract}

\maketitle

{\bf PACS (2003) classification:} 02.30 Xx, 02.30 Jr, 04.40 Nr

\bigskip

\section{Introduction}

The question whether the matter content of the universe is
accounted for by luminous objects has led to the search for mechanisms that
would lead to massive objects of a new kind. For instance, one
could consider the Einstein field equations with matter terms
arising from a real or complex scalar field that solves the (linear)
Klein-Gordon equation \cite{K,RB,SM}. Numerical computations
\cite{SS} suggest that such a configuration, with a real scalar field, should admit
long-lived, nearly periodic and strongly localized solutions. They
would owe their existence to the nonlinear structure of the
Einstein field equations, without any periodic forcing. In this
sense, they would be similar to the breather solution of the
sine-Gordon equation \cite{NLW,AS,L}. For this reason, objects
modeled by a solution of the Einstein field equations coupled to a
real Klein-Gordon field were called ``oscillating soliton stars,'' even though
they may not emit light, but merely affect light-rays in their
vicinity. Possible astrophysical applications, in particular
to dark matter, are discussed in \cite{SS}. If the scalar field decays very fast, it is expected
that the metric behaves like a Schwarzschild metric at infinity, with
mass related to the energy density of the scalar field. The
purpose of this paper is to provide a perturbative construction of
such soliton stars, thereby providing a simple set-up to
understand such objects analytically. We introduce two
assumptions:
\begin{itemize}
\item[(a)] the expansion parameter is an amplitude
parameter;
\item[(b)] the space and time variables are scaled in a way consistent
with the dispersion relation of the Klein-Gordon equation.
\end{itemize}
This procedure leads to a consistent limit, in which the equation
for the scalar field reduces to simple harmonic motion
\[ u_{tt}+u=0; \]
the amplitude depends on space, and is determined by a nonlinear
non-secularity condition at second order in perturbation theory
\cite{SK,NLW}. This procedure makes no reference to the complete
integrability of the sine-Gordon equation. This ``breather limit''
is also useful in particle physics \cite{DHN}. While the main
point of \cite{SK} was to give a rigorous argument to explain the
special role of the sine-Gordon equation in this context, the
method of proof also automatically generates approximate solitons,
by a procedure of general applicability. This paper gives an
analogue of this ``breather limit'' for the soliton star problem
and shows that the Einstein-Klein-Gordon equations admit a
consistent limit that is analytically tractable. It is the
counterpart of the Newtonian limit of the boson star problem with
complex scalar field \cite[(2.17--18)]{SS2}.

After recalling the field equations in Section 2, scaled variables
are introduced in Section 3 leading to the breather limit (Theorem
\ref{th:f}). At this stage, the metric and scalar field are
determined by solving a nonlinear coupled system in two unknowns
$S$ and $Z$. Intuitively, $Z$ is the Newtonian potential generated
by a mass density proportional to $S^2$, and $S$ solves a
Helmholtz equation with potential proportional to $Z$. A solution
for this nonlinear system is obtained in Section 4 (Theorem
\ref{th:v}, proved in Section 6), by a variant of the variational
method for finding nonlinear ``ground states'' of nonlinear
Klein-Gordon equations \cite{Strauss,BL}. The asymptotic behavior
of the metric is given in Theorem \ref{th:a}. In particular, the
metric component $g_{00}$ is the sum of a Schwarzschild-like term
and a periodic, exponentially decaying correction. Model
validation issues, using observations or computations, are
discussed in Section 5. It follows that light originating in the
vicinity of the soliton star should, if this model is valid,
exhibit a time-periodic frequency shift.

\section{Field equations}

Consider a spherically symmetric metric
\begin{equation}\label{eq:metric}
  ds^2=-N^2 dt^2+h^2 dr^2+r^2(d\theta^2+\sin^2\theta\; d\varphi^2),
\end{equation}
where $N$, $h$ and $\phi$ only depend on $(r,t)$. In the
following, subscripts $r$ and $t$ indicate derivatives with
respect to these variables. We let
$(x^0,\dots,x^3)=(t,r,\theta,\varphi)$, so that
$ds^2=g_{ab}dx^adx^b$, where Latin indices run from 0 to 3 and the
summation convention is used. The scalar field $\phi$ satisfies
the Klein-Gordon equation $-(-g)^{-1/2}\pa_a[g^{ab}(-g)^{1/2}\pa_b
\phi]+m^2\phi=0$, where $m$ is a positive constant. It derives
from the Lagrangian
\begin{equation} \label{eq:lagrm}
L=\frac{1}{2}(g^{ab}\pa_a\phi\pa_b\phi+m^2\phi^2)\sqrt {-g}.
\end{equation}
We assume that the metric and scalar field are even and periodic with respect
to $t$. We require $N\to 1$, $h\to 1$ and
$\phi\to 0$ as $r\to \infty$ for fixed $t$. The equations for the
scalar field and the metric coefficients are as follows \cite{SS}.
\begin{gather}\label{eq:scalar}
  -\frac{1}{N^2}
  \left[
    \phi_{tt}-\frac{N_t\phi_t}{N}+\frac{h_t\phi_t}{h}
  \right]
  +\frac{1}{h^2}
  \left[
    \phi_{rr}-\frac{h_r\phi_r}{h}+\frac{N_r\phi_r}{N}+\frac{2\phi_r}{r}
  \right]
=m^2\phi;
\\
\label{eq:rr}
  (N^2)_r=N^2
  \left[
  (h^2-1)/r+4\pi Grh^2(\frac{\phi_t^2}{N^2}+\frac{\phi_r^2}{h^2}
                       -m^2\phi^2)
  \right];
\\
\label{eq:00}
  (h^2)_r=h^2
  \left[
  -(h^2-1)/r+4\pi Grh^2(\frac{\phi_t^2}{N^2}+\frac{\phi_r^2}{h^2}
                       +m^2\phi^2)
  \right];
\\
\label{eq:0r}
  (h^2)_t=8\pi Grh^2\phi_r\phi_t.
\end{gather}

\section{Small-amplitude equations}

In this section, we define  new variables and scaled unknowns, and
prove that the field equations reduce, to leading order, to a system of two equations in two
unknowns.

\subsection{New variables and unknowns} Define new variables by
\begin{eqnarray*}
\xi&=&m\ep r,\\ \tau &=&mt\sqrt{1-\ep^2},
\end{eqnarray*}
where $\ep>0$ is a new parameter. The scaling of time variables is
motivated by the form of the Klein-Gordon equation.\footnote{The
function $f=\exp(\ep mr)\cos(\omega t)$ solves
$f_{tt}-f_{rr}+m^2f=0$ if and only if $\omega^2=m^2(1-\ep^2)$.}
Derivatives transform as follows:
\[ \pa_r=\ep m\pa_\xi;\quad \pa_t=m(1-\ep^2)^{1/2}\pa_\tau.
\]
In the following, subscripts $\xi$ and $\tau$ denote derivatives with respect to
these variables. We let
\[\mu=4\pi G.\]

Next, define new unknowns $u$, $\tilde N$ and $\tilde h$ by
\begin{subeqnarray}\label{eq:Ndef}
\phi&=&\ep^2u(\xi, \tau,\ep),\\ N^2 &=&1+\ep^2 \tilde N(\xi, \tau,\ep),\\ h^2
&=&1+\ep^2 \tilde h(\xi, \tau,\ep).
\end{subeqnarray}
The assumptions on the metric and scalar field in section 2 lead to the conditions
\begin{equation*}\tag{7d}
 u,\;\tilde N {\rm\ and\ } \tilde h {\rm\ tend\ to\ } 0{\rm\ as\  }r\to\infty,
 \text{ and are } 2\pi\text{-periodic in }\tau.
\end{equation*}
As a result, the field variables are $t$-periodic with period
$2\pi/\omega$, with $\omega=m(1-\ep^2)^{1/2}$. While, for the linear Klein-Gordon
equation, the amplitude of the solution may be chosen
independently of the period, this is not true in the nonlinear
case, since $\phi$ is determined by $\ep$, and therefore by
$\omega$. A general feature of nonlinear oscillators is the dependence of
amplitude on period; here, this dependence is reflected in equations (9).
In particular, $\omega$ determines
the leading order amplitude $\ep^2 S(\xi)$ of the scalar field $\phi$.

\subsection{Field equations at leading order}
We prove the following result.
\begin{theorem}\label{th:f}
To lowest order in $\ep$, the soliton star problem
(\ref{eq:scalar}--\ref{eq:Ndef}) is equivalent
to the system
\begin{subeqnarray}\label{eq:sz}
\Delta S-S &=&SZ; \\
\Delta Z   &=& \mu S^2;\\
S\to 0 {\rm\ and\ } Z\to 0 & {\rm as} & \xi\to\infty.
\end{subeqnarray}
where $\Delta=d^2/d\xi^2+(2/\xi)d/d\xi$ is the radial
Laplacian in three dimensions and $\mu=4\pi G$.
The functions $S$ and $Z$ determine the scalar field and the
metric at order $\ep^2$ via
\begin{subeqnarray}\label{eq:umh}
\phi &=& \ep^2 S(\xi)\cos \tau;\\ N^2  &=& 1+\ep^2 (Z+Y\cos
2\tau);\\  h^2  &=& 1+\ep^2 h_0(\xi),
\end{subeqnarray}
where
\begin{equation}\label{eq:h0}
h_0(\xi)=\frac{\mu}{\xi}\int_0^\xi y^2 S^2(y)dy
\quad
{\rm and}\quad
Y(\xi)=\int_\xi^\infty \mu y S^2(y)dy.
\end{equation}
\end{theorem}
\begin{remark}
The $1/\xi$ decay of $h^2-1$ is consistent with the behavior of
the Schwarzschild solution. The integral for $Y$ converges because
the scalar field decays exponentially fast at infinity, by Theorem
\ref{th:a}.
\end{remark}
\begin{proof}
Write
\[u=u_0+\ep^2u_1+\dots,\quad
\tilde N=N_0+\ep^2N_1+\dots,\quad \tilde h=h_0+\ep^2h_1+\dots,
\]
and insert into equations (\ref{eq:scalar}--\ref{eq:0r}).
In the following, we lump as $\O(\ep^k)$ various terms which
involve a power of $\ep$ equal to $k$ or more.
Equations (\ref{eq:rr}), (\ref{eq:00}) and (\ref{eq:0r}) become respectively
\begin{subeqnarray}
\tilde N_\xi &=&\frac {\tilde h}\xi+\mu\xi(u_\tau^2-u^2)+\O(\ep^2);\\
\tilde h_\xi &=& -\frac {\tilde h}\xi+\mu\xi(u_\tau^2+u^2)+\O(\ep^2);\\
\tilde h_\tau& =& \O(\ep^2).
\end{subeqnarray}
Using the relations (\ref{eq:Ndef}), equation (\ref{eq:scalar})
becomes
\begin{equation}
\label{eq:this} u
=-\frac{(1-\ep^2)}{1+\ep^2\tilde N}\left[u_{\tau\tau}-\frac12\ep^2(\tilde N_\tau-\tilde
h_\tau)u_\tau\right]
+\ep^2\frac{u_{\xi\xi}+2u_\xi/\xi}{1+\ep^2
h}+\O(\ep^4).
\end{equation}
Let
\[ \Delta =
\frac{\pa^2}{\pa \xi^2}+\frac 2\xi\frac\pa{\pa\xi}.
\]
Since $(1-\ep^2)/(1+\ep^2\tilde N)=1-\ep^2(1+\tilde N)+\O(\ep^4)$, (\ref{eq:this}) may be simplified to
\begin{equation}\label{eq:d}
u_{\tau\tau}+u=
\ep^2\left\{
(1+\tilde N)u_{\tau\tau}-\frac12(\tilde N_\tau-\tilde h_\tau)u_\tau+\Delta u
\right\} +\O(\ep^4).
\end{equation}

At leading order, we therefore obtain the equations
\begin{subeqnarray}
h_{0\tau} &=& 0, \label{eq:0a}\\
h_{0\xi}  &=&-\frac {h_0}\xi+\mu\xi(u_{0\tau}^2+u_0^2),\\
N_{0\xi}  &= &\frac {h_0}\xi+\mu\xi(u_{0\tau}^2-u_0^2),\\
u_{0\tau\tau}+u_0 &=& 0.
\end{subeqnarray}
Equation (\ref{eq:0a}d) gives, since $u$ is even in time,
\[ u_0=S(\xi)\cos\tau.
\]
This proves (\ref{eq:umh}a).
It follows that $u_{0\tau}^2+u_0^2=S^2$; equation (\ref{eq:0a}b)
now yields
\begin{equation}\label{eq:21}
h_{0\xi}  =-\frac {h_0}\xi+\mu\xi S^2,
\end{equation}
and (\ref{eq:0a}a) shows that
\[ h_0=h_0(\xi),
\]
hence (\ref{eq:umh}c).
Finally, (\ref{eq:0a}c) gives
\begin{equation}\label{eq:N0}
N_{0\xi}=\frac {h_0(\xi)}\xi-\mu\xi S^2\cos 2\tau.
\end{equation}
It follows that $N_0 = Y(\xi)\cos2\tau+Z(\xi) +N_{00}(\tau)$.
Since the metric is asymptotically flat as $\xi\to \infty$, $N_{00}$ is constant.
Incorporating it into $Z$, we obtain
\begin{equation}
\label{eq:m0}
N_0 = Z(\xi)+Y(\xi)\cos2\tau,
\end{equation}
where $Y$ and $Z$ tend to zero as $\xi\to\infty$.
Equation (\ref{eq:umh}b) therefore holds.

Equation (\ref{eq:N0}) now yields
\begin{eqnarray}
Z_\xi &= & h_0(\xi)/\xi ,\label{eq:Z16}\\
Y_\xi & =& -\mu\xi S^2. \label{eq:Y16}
\end{eqnarray}
Equations (\ref{eq:21}) and (\ref{eq:Z16}) now yield
\[ \mu\xi^2S^2=(\xi h_0)_\xi=(\xi^2 Z_\xi)_\xi=\xi^2 (Z_{\xi\xi}+\frac 2\xi Z_\xi) =\xi(\xi Z)_{\xi\xi},
\] hence
\[ \Delta Z = \mu S^2.
\]
This proves (\ref{eq:sz}a).
Finally, equation (\ref{eq:0a}b) yields $(\xi h_0)_\xi=\mu\xi^2 S^2$;
since $h_0$ should be regular at the origin, the first part of
(\ref{eq:h0}) follows. Equation  (\ref{eq:Y16}) gives the rest of
(\ref{eq:h0}). The convergence of the integral is a consequence of
the exponential decay of $S$ (see Theorem \ref{th:a}).

It remains to prove (\ref{eq:sz}b). To this end, consider
the terms of order $\ep^2$ in equation
(\ref{eq:d}):
\begin{equation}
\label{eq:1d}
u_{1\tau\tau}+u_1 = (1+N_0)u_{0\tau\tau}+\frac12 N_{0\tau}u_{0\tau}
+\Delta S\cos\tau.
\end{equation}
Now,
\begin{eqnarray*}
\lefteqn{(1+N_0)u_{0\tau\tau}+\frac12 N_{0\tau}u_{0\tau}  } & &\\
& = & -(1+Z+Y\cos 2\tau)S\cos\tau+(Y\sin 2\tau)(S\sin \tau)\\
& = & -(1+Z)S\cos\tau-YS\cos 3\tau.
\end{eqnarray*}
Since $u_1$ should be $2\pi$-periodic in
$\tau$, the right-hand side of (\ref{eq:1d})
should not contain any term proportional to $\cos\tau$. Therefore,
\[\Delta S-(1+Z)S=0.
\]
This completes the proof of the theorem.
\end{proof}
\begin{remark}
The function $Y$ may also be expressed directly in terms of $Z$:
since $Y_\xi=-\mu\xi S^2$, $(Y+(\xi Z)_\xi)_\xi=0$ hence, since
$Y$, $Z$ and $\xi Z_\xi= h_0$ all tend to zero as $\xi\to\infty$,
\[ Y=-(\xi Z)_\xi.
\]
\end{remark}

\section{Solutions of system (\ref{eq:sz})}

We now introduce a variational formulation of system (\ref{eq:sz})
and show that minimizing sequences converge to a solution
in which $S$ decays exponentially and $Z$ behaves
like a Newtonian potential at infinity.

\subsection{Variational principle and existence of a solution}

Define
\begin{equation}\label{eq:e}
E[S,Z] = \frac{1}{2}\int_0^\infty [S^{\prime 2}+S^2 + Z^{\prime
2}/(2\mu)] \xi^2 d\xi,
\end{equation}
where the prime denotes the derivative with respect to $\xi$, and
\begin{equation}\label{eq:i}
I[S,Z] = \frac12\int_0^\infty S^2Z \xi^2 d\xi.
\end{equation}
These expressions may be written
\[ E = \frac1{8\pi}\int_{\RR^3}[|\nabla S|^2+S^2+\frac{|\nabla Z|^2}{2\mu}]d^3x
\]
where $\nabla$ denotes the (Euclidean) gradient in $\RR^3$,
and
\[ I = \frac1{8\pi}\int_{\RR^3}S^2Zd^3x.
\]
System (\ref{eq:sz}) is the Euler-Lagrange equation of the
Lagrangian $F[S,Z]:=E[S,Z]+I[S,Z]$. This expression is
unbounded below.\footnote{Choose $S$ and $Z$ so that $I[S,Z]<0$;
then $F[aS,aZ]\to-\infty$ as $a\to+\infty$.}
We therefore minimize $E$ while keeping the value of $I$ fixed. We may
assume $I=-1$ without loss of generality: since $I$ is homogeneous
of degree 3, its value may be modified by scaling as long as it
is nonzero.\footnote{The infimum of $E$ over the set of functions
such that $I=0$ is clearly zero.} We let $(S,Z)$ vary over the
space $H^1_r(\RR^3)\times D^1_r(\RR^3)$, where $H^1_r(\RR^3)$ is the space of
radial functions $u(\xi)$ on $\RR^3$ that are square-summable
together with their first-order derivatives, while
$D^1_r(\RR^3)$ is the closure of the set of compactly supported,
smooth radial functions, for the norm $\norm{\nabla u}_{L^2}$;
$D^1_r(\RR^3)$ may also be viewed as the space of radial functions in $L^6(\RR^3)$
that have square-summable first-order derivatives \cite{BL}. The
space  $H^1_r(\RR^3)\times D^1_r(\RR^3)$ is a Hilbert space with
pairing
\[ \scal{S_1,Z_1}{S_2,Z_2}=
\int[\nabla S_1\cdot\nabla S_2 + S_1S_2 + \nabla Z_1\cdot\nabla
Z_2]d^3x,
\]
and norm
\[ \norm{S,Z}=\sqrt{\scal{S,Z}{S,Z}}.
\]

We prove three results:
\begin{itemize}
\item The infimum of $E$ constrained by $I$ is achieved for some $(S,Z)$.
\item $(S,Z)$ generate a solution of (\ref{eq:sz}).
\item At infinity, the metric is Schwarzschild-like, and $S$
decays exponentially.
\end{itemize}
The first two points follow from the following theorem. The third
is proved in section \ref{sec:a}.

\begin{theorem}\label{th:v}
The infimum of $E$ subject to the constraint $I=-1$, where $(S,Z)$
varies in $H^1_r(\RR^3)\times D^1_r(\RR^3)$, is achieved; after
scaling $Z$, this provides a solution of (\ref{eq:sz}).
Futhermore, any solution of (\ref{eq:sz}) in this function space satisfies
\begin{equation}
\label{eq:estM}
M := \mu \int_0^\infty \xi^2S(\xi)^2d\xi\geq  \frac38\pi\sqrt
3.
\end{equation}\end{theorem}
The somewhat technical proof is deferred to section \ref{sec:pf}.

\begin{remark}
The Lagrangian $F$ may be given a geometric interpretation.
Consider $\mathcal L[N,h,\phi]\sin\theta \,dr\,d\theta\,d\varphi
=(R\sqrt{-g}-2\mu L)\sin\theta \,dr\,d\theta\,d\varphi$, where
$L$ is the Lagrangian density (\ref{eq:lagrm}) for the scalar
field, and $R\sqrt{-g}$ is the Hilbert Lagrangian. Next, perform
the change of unknown (\ref{eq:Ndef}): this leads to the
expression
\[ \mathcal M[S,Z,\ep,\tau] = \mathcal L[1+\ep^2(Z+Y\cos\tau),1+\ep^2h_0,\ep^2 S\cos
\tau].
\]
Expand $\mathcal M$ with respect to $\ep$: $\mathcal M = \mathcal
M_0+\ep^2\mathcal M_2+\ep ^4\mathcal M_4+\cdots$ Then, take the average
of each term with respect to $\tau$ over one period:
$\langle\mathcal M\rangle=\langle\mathcal
M_0\rangle+\ep^2\langle\mathcal M_2\rangle+\cdots$ After
computation, one obtains $\langle \mathcal M_0\rangle=0$, $\langle
\mathcal M_2\rangle=(\xi Z)_\xi$, and $\langle \mathcal
M_4\rangle=-\frac14\xi^2\{(2\mu)(S^{\prime 2}+S^2(1+Z))+Z^{\prime
2}\}+\Phi_\xi$, where $\Phi$ is a function of the field variables
and their derivatives. Since $\Phi_\xi$ and $(\xi Z)_\xi$ are
divergences, they do not contribute to the Euler equation. We are
left with $\langle \mathcal M_4\rangle$, which differs from $F$ by
a multiplicative constant.
\end{remark}

\subsection{Decay estimates}
\label{sec:a}

We now turn to the decay properties of $S$ and $Z$. We state the
result using the original variable $r$ so that the result should
be easier to interpret.
\begin{theorem}\label{th:a}
For any $\alpha < 1$, $S=\O(e^{-\alpha \ep m r})$ and
$r Z(\xi)\to -M/(\ep m)$ as $r\to\infty$, where
\begin{equation}
M =\mu\int_0^\infty y^2S^2(y)dy.
\end{equation}
In addition, $S\sqrt\mu$ is independent of $\mu$.
\end{theorem}
Thus, the metric is Schwarzschild-like:
\begin{equation} \label{eq:dec}
N^2\approx 1-\frac{\ep M}{m r},\quad\text{and}\quad
h^2\approx 1+\frac{\ep M}{m r}.
\end{equation}
\begin{proof}
The argument is classical \cite{Strauss,BL}.
System (\ref{eq:sz}) may be written
\begin{subeqnarray}\label{eq:simpl}
(\xi S)'' &=&\xi S(Z+1); \\  (\xi Z)'' &= &\mu \xi S^2,
\end{subeqnarray}
Therefore, $S=S_1\sqrt\mu$, where $S_1$ solves the same system
with $\mu=1$. The decay of higher derivatives follows
from interior regularity estimates for the Laplacian; it therefore suffices to
prove the decay of $S$ and $Z$.

Let $v(\xi)=\xi S$ and $w(\xi)=\frac12 v^2$. Since $Z\to 0$ at
infinity, one may, for any $\alpha\in (0,1)$, find $R>0$ such that
$1+Z(\xi)\geq\alpha$ for $\xi\geq R$. Let $\beta=\sqrt{2\alpha}$
and $\psi(\xi)=(w'+\beta w)e^{-\beta\xi}$, so that
\[ e^{\beta\xi}\psi'= w''-\beta^2 w = vv''+v^{\prime\, 2}-\alpha
v^2=[(1+Z)-\alpha]v^2+v^{\prime\, 2}\geq 0,
\]
so that $\psi$ is nondecreasing for $\xi\geq R$. There are now two possibilities.
\begin{itemize}
\item[(a)]
For every $\xi\geq R$, $\psi(\xi)\leq 0$. In that case, for $\xi\geq R$, we have
$(we^{\beta\xi})'=(w'+\beta w)e^{\beta\xi}\leq 0$, hence
\[ w\leq w(\xi)\leq w(R)e^{-\beta\xi}.
\]
This proves that $w$, hence $S$, decays exponentially.

\item[(b)]
There is a $\xi_0\geq R$ such that $\psi(\xi_0)> 0$. Since $\psi$
is nondecreasing, $w'(\xi)+\beta
w(\xi)\geq\psi(\xi_0)e^{\beta\xi}$ for $\xi\geq \xi_0$, and
$w'+\beta w$ is not integrable near infinity. However,
$\int_0^\infty w\,d\xi=\int_0^\infty
\frac12\xi^2S^2\,d\xi=(8\pi)^{-1}\norm{S}_{L^2}^2<\infty$, and
$|w'|=|\xi S(\xi S'+S)|\leq\frac12\xi^2(S^2+S^{\prime 2})+\xi S^2$
is also integrable near infinity; indeed,
$\int_0^\infty\xi^2S^{\prime 2}\,d\xi$ equals
$(4\pi)^{-1}\norm{\nabla S}_{L^2}$, and is therefore finite. This
contradiction proves that case (b) cannot occur.
\end{itemize}
This completes the proof of the exponential decay of $S$.

Equation (\ref{eq:simpl}b) for $Z$ may now be integrated: since
$\xi Z$ vanishes for $\xi=0$,
\[
Z=z_0+\frac{\mu}{\xi}\int_0^\xi (\xi-y)yS^2(y)dy
\]
where $z_0$ is constant. Since $S$ decays exponentially,
$Z$ has a limit at infinity. This limit must be zero,
since $Z\in L^6$. Therefore,
$z_0=-\mu\int_0^\infty yS^2(y)dy$, and
\begin{eqnarray*}
Z(\xi) & = & -\mu\int_\xi^\infty yS^2(y)dy-\frac{\mu}{\xi}\int_0^\xi
y^2S^2(y)dy\\
&=&-\mu\int_\xi^\infty yS^2(y)dy-
\frac1{\xi}\left[M-\mu\int_\xi^\infty y^2S^2(y)\right]\\
&=& -\frac M\xi +\int_\xi^\infty (\frac y\xi -1)\mu y S^2(y)dy.
\end{eqnarray*}
Equation (\ref{eq:21}) gives the behavior of $h_0$. The behavior
of $N$ and $h$ follow.
This completes the proof.
\end{proof}
\begin{remark}
Since Poisson's equation admits singular solutions,
corresponding to point masses, it is natural to ask whether system
(\ref{eq:sz}) admits solutions in which $Z$ behaves like $1/\xi$
at the origin. A positive answer may be obtained by the method of
Reduction : following the strategy described in \cite{FR}, one can
prove that there is a four-parameter family of solutions, defined
for small $\xi$, such that
\begin{eqnarray*}
S(\xi) &=&\frac{S_0}\xi+S_1+S_{11}\ln\xi+\O(\xi\ln\xi);\\
Z(\xi) &=&\frac{Z_0}\xi+Z_1+Z_{11}\ln\xi+\O(\xi\ln\xi).
\end{eqnarray*}
where $S_0$, $S_1$, $Z_0$, $Z_1$ are arbitrary constants,
$S_{11}=S_0Z_0$, and $Z_{11}=\mu S_0^2$.
The solutions considered so far all have $S_0=Z_0=0$. A
consequence of this computation is that it is possible to take
$S_0=0$ and $Z_0\neq 0$: a singularity in $Z$ does not necessarily
imply a singularity in the scalar field.
There is also a
reduction with $S$ and $Z$ behaving like $1/\xi^2$, but it
involves fewer constants.
Note also that system (\ref{eq:sz}) admits a one-parameter family
of scaling transformations:
\[(S,Z,\theta)\mapsto(\theta^2 S(\theta \xi),\theta^2 Z(\theta
\xi)+\theta^2-1).
\]
\end{remark}

\section{Model validation}
\label{sec:r}

The information on the asymptotic behavior of $S$ and $Z$ now
enables us to relate the parameters $\ep$ and $m$ of the model to
observational data. Since we need two parameters, we need two
data. For instance, consider the redshift of light originating at
two points $B$ and $B'$, of known location ($r=r_B$ and $r_{B'}$)
and observed at a point $A$ relatively at rest with respect to $B$
and $B'$ (see \cite{Ksr} for the precise meaning of relative
velocity in general relativity for distant objects). We obtain
\[ \frac{\nu_B-\nu_A}{\nu_A} =\frac{(g_{00})_B}{(g_{00})_A}-1\approx
     Z(\ep m r_B)+S(\ep m r_B)\cos (mt\sqrt{1-\ep^2}),
\]
where subscrpts $A$ and $B$ indicate the points where the
frequency ($\nu$) and metric components ($g_{00}$) are determined.
If $r_B$ is large, $S(\xi_B)$ is small relative to
$Z(\xi_B)\approx 1-\frac{\ep M}{mr_B}$, because of the fast decay
of $S$. The redshift for this source should therefore be nearly
constant in time. From it, we may estimate $\ep/(mr_B)$, hence the
parameter $f=\ep/m$.

Next, consider a source $B'$ closer to the
center of the putative soliton star. If the present model is
correct, one should now observe an \emph{oscillatory redshift}, with
period $T=2\pi/(m\sqrt{1-\ep^2})$. Eliminating $m$ gives
\begin{equation} \label{eq:fT}
\frac {2\pi f} T = \ep\sqrt{1-\ep^2};
\end{equation}
since $\ep\in (0,1)$, this expression can never exceed $\frac 12$.
Thus, from the observation of $f$ and $T$, we may check whether
$2\pi f/T$ is less than $1/2$ and, if that is the case, compute
$\ep$ from (\ref{eq:fT}), and deduce the value of $m$ from the
relation $m=\ep/f$. Since  (\ref{eq:fT}) has in general two roots,
the smaller one seems preferable, in view of the assumption of
small amplitude.

We may also estimate parameters from the numerical data in
\cite{SS,SS2}, where the total mass, defined as
\[ \lim_{r\to\infty} \frac r2 (1-h^{-2}),
\]
is equal to $0.52$, and the period is $2\pi/\omega$ with $\omega=0.0196$. In the
present notation, this means: $\ep M/m=0.52$,
$m\ep\sqrt{1-\ep^2}=0.0196$. Therefore, $\ep^2\lessapprox 0.52\times 0.0196/M$, hence, using
estimate (\ref{eq:estM}),
\[ \ep\lessapprox 5 \times 10^{-3}.
\]

\section{Proof of Theorem \ref{th:v}}
\label{sec:pf}

\subsection{Step 1 : Convergence of minimizing sequences}

Consider a minimizing sequence $(S_n,Z_n)$:
\[ E[S_n,Z_n] \text{ decreases and tends to } E_0 := \inf_{I(S,Z)=-1} E[S,Z]
\]
as $n$ tends to infinity.
Since $E[S_n,Z_n]$ is in particular bounded,
one can prove\footnote{For the results on function spaces used here
see \cite{Strauss,BL}. In addition, functions $S$
in $H^1_r(\RR^3)$ satisfy an estimate of the form
\[ |S(\xi)|\leq c_2 \xi^{-1}(\|\nabla S\|_{L^2}+\|S\|_{L^2}).
\]
Functions $Z$ in $D^1_r(\RR^3)$ satisfy an estimate of the form
\[ |Z(\xi)|\leq c_3 \xi^{-1/2}\|\nabla Z\|_{L^2}.
\]
In particular, if $u$ belongs to  $D^1_r$, it satisfies $\xi u(\xi)\to
0$ as $\xi\to 0$. For background results on weak convergence, see \cite{B}.}
that for any $p\in(2,6)$, any
bounded sequence in $H^1_r(\RR^3)$ admits a subsequence that
converges weakly in $H^1_r(\RR^3)$, strongly in $L^p(\RR^3)$, and
pointwise almost everywhere. In other words, there is a sequence,
still called $(S_n, Z_n)$ for convenience, and a pair
$(S,Z)\in H^1_r(\RR^3)\times D^1_r(\RR^3)$ such that the following
properties hold simultaneously
\begin{gather}
I[S_n,Z_n]=-1 \text{ for every } n\text{ and } \lim_{n\to\infty} E[S_n,Z_n] = E_0,\\
(\forall (\sigma,\zeta)\in H^1_r\times D^1_r)  \quad \lim_{n\to\infty}\scal{S_n,Z_n}{\sigma,\zeta}= 0,\\
\lim_{n\to\infty}\norm{S_n - S}_{L^{p}} = 0,\\
S_n \to S \quad\text{ almost everywhere},\\
(\forall \zeta\in L^{6/5})  \quad \lim_{n\to\infty}\int{Z_n}{\zeta}d^3x = 0.
\label{eq:wk}
\end{gather}

\subsection{Step 2 : $(S,Z)$ is a minimizer}

We need to prove that $E[S,Z]=E_0$ and $I[S,Z]=-1$.
Since the norm in any Hilbert space is weakly lower semi-continuous,
\[ E[S,Z]\leq E_0=\liminf_{n\to\infty} E[S_n,Z_n],
\]
and since $E[S,Z]$ cannot be lower than its infimum $E_0$, we conclude
\[ E[S,Z]=E_0.
\]
Next, let us choose $p=12/5$. Since $\|S_n- S\|_{L^{12/5}}\to 0$,
$S$ and $S_n$ are both bounded in $L^{12/5}$, and therefore, using
H\"older's inequality, which applies since $5/12+5/12+1/6=1$,
\[ \left|\int_{\RR^3}(S_n^2-S^2)Z_nd^3x\right| \leq
\|S_n+ S\|_{L^{12/5}}\|S_n- S\|_{L^{12/5}}\|Z_n\|_{L^{6}}\to 0
\]
as $n$ tends to infinity. On the other hand, since $S\in L^{12/5}$,
$S^2\in L^{6/5}$. Using (\ref{eq:wk}), we obtain
\[ \lim_{n\to\infty}\int_{\RR^3}S^2(Z_n-Z)d^3x = 0.
\]
Writing $S_n^2Z_n-S^2Z=(S_n^2-S^2)Z_n+S^2(Z_n-Z)$, we obtain
\[ \lim_{n\to\infty}I[S_n,Z_n]=I[S,Z],
\]
hence $I[S,Z]=-1$.

\subsection{Step 3 : $E$ and $I$ are of class $C^1$ on
$H^1_r(\RR^3)\times D^1_r(\RR^3)$}

This is true for $E$ because
it is the square of the norm in a Hilbert space. As for $I$, its
G\^ateaux derivative is the map $dI$ defined by
\[ dI[S,Z]\cdot(\sigma,\zeta)=\int(SZ\sigma+\frac12S^2\zeta)d^3x.
\]
A form of the Sobolev embedding theorem \cite{Strauss,BL} shows
that there is a constant $c_4$ such that
\[ \norm{S}_{L^{12/5}} + \norm Z_{L^6}\leq c_4 \norm{S,Z}.
\]
Since $5/12+1/6=7/12$ and $5/12+5/12=5/6$, H\"older's inequality
yields
\[ \norm{SZ}_{L^{12/7}} + \norm {S^2}_{L^{6/5}}\leq c_5 \norm{S,Z}^2.
\]
Since $\int(SZ\sigma+\frac12S^2\zeta)d^3x\leq
\norm{SZ}_{L^{12/7}}\norm\sigma_{L^{12/5}}+\norm
{S^2}_{L^{6/5}}\norm {\zeta}_{L^{6}}$,
\[ |dI[S,Z]\cdot(\sigma,\zeta)|\leq c_6 \norm{S,Z}^2
\norm{\sigma,\zeta}.
\]
This proves that $dI$ is a continuous linear form on $H_r^1\times
D^1_r$.

Regarding the continuity of $dI$ with respect to $(S,Z)$,
\begin{eqnarray*}
\lefteqn{|(dI[S_1,Z_1]-dI[S_2,Z_2])\cdot(\sigma,\zeta)| } & & \\
 & = & \int [(S_1-S_2)Z_1+S_2(Z_1-Z_2)\sigma
 + \frac12(S_1-S_2)(S_1+S_2)\zeta]d^3x \\
 & \leq &  \left\{\norm{S_1-S_2}_{L^{12/5}}\norm{Z_1}_{L^6}
 +\norm{S_2}_{L^{12/5}}\norm{Z_1-Z_2}_{L^6}\right\}\norm{\sigma}_{L^{12/5}}
\\
& & \mbox{\quad} +
\frac12\norm{S_1-S_2}_{L^{12/5}}\norm{S_1+S_2}_{L^{12/5}}\norm{\zeta}_{L^6}
\\
& \leq &
c_7\norm{\sigma,\zeta}(\norm{S_1,Z_1}+\norm{S_2,Z_2})\norm{S_1-S_2,Z_1-Z_2}.
\end{eqnarray*}
The continuity of $dI$ follows.

\subsection{Step 4 : $(S,Z)$ solves (\ref{eq:sz})}

We first prove that
there is a Lagrange multiplier $\lambda$ such that $dE=\lambda dI$
or, in other words, that
\begin{equation} \label{eq:syssz}
 -\Delta S + S =\lambda SZ;\quad -\Delta Z = \lambda\mu S^2.
\end{equation}
Because of the constraint, $S$ is not identically zero; therefore,
we may a find a positive function $Z_0$ with compact support such that
\[ \int S^2Z_0d^3x=1.
\]
If $\sigma$ and $\zeta$ are arbitrary variations of $S$ and $Z$,
so that $I[S+\sigma,Z+\zeta]$ may not be equal to $-1$, one may,
if $\sigma$ is small enough in $D^1_r$,\footnote{This smallness
condition guarantees that $\int (S+\sigma)^2Z_0d^3x\neq 0$.}
define a constant $\theta$ by
\[ I[S+\sigma,Z+\zeta-\theta Z_0]=-1.
\]
The result is
\begin{eqnarray*}
\theta=\theta(S,Z,\sigma,\zeta)
 & = &\frac{\int[(S+\sigma)^2\zeta+(2S+\sigma)Z\sigma]d^3x}{\int (S+\sigma)^2Z_0 d^3x}
\end{eqnarray*}
One then expresses the stationarity condition:
\[ \frac{\mathrm d}{{\mathrm d}t}E[S+t\sigma,Z+t\zeta-\theta(S,Z,t\sigma,t\zeta)Z_0]=0
\]
for $t=0$. Since ${\mathrm d}\theta/{{\mathrm d}
t}=\int[S^2\zeta+2SZ\sigma]d^3x$ for $t=0$, we obtain the
condition
\[ \int[\nabla S\cdot\nabla\sigma+S(1-2\lambda Z)\sigma+\frac{\nabla Z}{2\mu}\nabla\zeta
-\frac\lambda 2 Z S^2\zeta]d^3x=0,
\]
where $\lambda=\int\nabla Z\cdot\nabla Z_0 d^3x$.
After integration by parts, this leads to (\ref{eq:syssz}).

If $\lambda$ were equal to zero, $Z$ would be a singularity-free
solution of Laplace's equation which tends to zero at infinity. By
Liouville's theorem, $Z$ would then be identically zero, hence
$I[S,Z]=0$, violating the constraint. Therefore, $\lambda\neq 0$.
This allows us to consider $(S/\lambda,-Z/\lambda)$, which solves
(\ref{eq:sz}). This completes the proof.
\begin{remark}
Multiplying (\ref{eq:sz}a) by $S$ and integrating by parts (the
boundary term which arises in this manner vanishes because $S$ and
its derivatives decay exponentially), we obtain
\[ \int_{\RR^3}[|\nabla S|^2+S^2]dx = 2\lambda I[S,Z]=-2\lambda.
\]
Therefore, $\lambda <0$.
\end{remark}

\subsection{Step 5: estimate (\ref{eq:estM}) holds}

Since $M=4\pi\mu \norm{S}_{L^2}^2$, it suffices to estimate the
$L^2$ norm of $S$. Multiplying (\ref{eq:sz}a)--(\ref{eq:sz}b) by
$S$ and $Z$ respectively, and integrating, we obtain
\begin{equation}
\label{eq:estsz}
\int(|\nabla S|^2+S^2)d^3x=-\int S^2Zd^3x=\frac 1\mu\int |\nabla
Z|^2d^3x.
\end{equation}
Using H\"older's inequality and interpolation, we obtain
\[\left| \int S^2Zd^3x \right| \leq
  \norm{Z}_{L^6}\norm{S^2}_{L^{6/5}} =
  \norm{Z}_{L^6}\norm{S}_{L^{12/5}}^2 \leq
  \norm{Z}_{L^6}\norm{S}_{L^2}^{3/2}\norm{S}_{L^6}^{1/2},
  \]
since $\frac 5{12} = \frac 34\times\frac12+\frac14\times\frac16$.
Let $K$ denote the best constant in Sobolev's inequality:
\[ \norm{u}_{L^6}\leq K  \norm{\nabla u}_{L^2}.
\]
It is known\footnote{See \cite[p.~39 sqq.]{aubin} and \cite{aubin2,talenti,rosen}.} that
\[ K = \frac 1{\sqrt3}\left( \frac 4 {\pi^2} \right)^{1/3}.
\]
Equation (\ref{eq:estsz}) implies
\[ K^{-2}\norm{Z}_{L^6}^2\leq
 \int |\nabla Z|^2d^3x \leq \mu
 \norm{Z}_{L^6}\norm{S}_{L^2}^{3/2}\norm{S}_{L^6}^{1/2},
\]
hence $\norm{Z}_{L^6}\leq \mu
K^2\norm{S}_{L^2}^{3/2}\norm{S}_{L^6}^{1/2}$, and
\begin{eqnarray*}
 K^{-2}\norm{S}_{L^6}^2+\norm{S}_{L^2}^2
 & \leq & \int(|\nabla S|^2+S^2)d^3x  \\
 & \leq &  \norm{Z}_{L^6}\norm{S}_{L^2}^{3/2}\norm{S}_{L^6}^{1/2}
 \\
 & \leq & \mu K^2 \norm{S}_{L^2}^{3}\norm{S}_{L^6}.
\end{eqnarray*}
Letting $X=\norm{S}_{L^6}$, we conclude that
\[ K^{-2}X^2-\mu K^2 \norm{S}_{L^2}^{3}X+\norm{S}_{L^2}^2\leq 0.
\]
The discriminant of this quadratic expression in $X$ must therefore be
nonnegative. This yields the relation $\mu^2
K^4\norm{S}_{L^2}^{6}-4K^{-2}\norm{S}_{L^2}^2\geq 0$, or
$\mu\norm{S}_{L^2}^2\geq 2K^{-3}$. It follows that
\[ M=\mu\norm{S}_{L^2}^2(4\pi)^{-1}\geq
2 \frac{\pi^2}43^{3/2}(4\pi)^{-1}=\frac 38 \pi\sqrt 3,
\]
QED.


\begin{thebibliography}{99}

\bibitem{AS}
M. Ablowitz and H. Segur,
\emph{Solitons and the Inverse Scattering Transform},
SIAM, Philadelphia, 1981.
\bibitem{aubin}
T. Aubin,
\emph{Nonlinear Analysis on Manifolds. Monge-Amp\`ere equations},
Springer, Berlin, 1982.
\bibitem{aubin2}
T. Aubin, Probl\`emes isop\'erim\'etriques et espaces de Sobolev,
J. Diff. Geom. {\bf 11} (1976) 573-598, C. R. Acad. Sci. Paris {\bf
280} A (1975) 279.
\bibitem{BL}
H. Berestycki and P.-L. Lions,
Nonlinear scalar field equations,
Arch. Rat. Mech. Anal. {\bf 82} (1983) 313--375.
\bibitem{B}
H. Brezis, \emph{Analyse Fonctionnelle : Th\'eorie et
Applications}, Masson, Paris 1983.
\bibitem{DHN}
 R. Dashen, B. Hasslacher, A. Neveu,
Particle spectrum in model field theories from semiclassical
functional integral techniques,
Phys. Rev. D, {\bf 11}, 1975, 3424--56.
\bibitem{K} D. J. Kaup, Klein-Gordon geon,
Phys. Rev. {\bf 172} : 5 (1968) 1331--1342.
\bibitem{Ksr} S. Kichenassamy, Sr.,
Compl\'ements \`a l'interpr\'etation physique de la Relativit\'e
g\'en\'erale: Applications,
Ann. Inst. H. Poincar\'e, section A, {\bf 1} : 2 (1964) 129--145.
\bibitem{SK} S. Kichenassamy,
Breather solutions of the nonlinear wave equation,
Comm.\ Pure and Appl.\ Math. {\bf 44} (1991) 789--818.
\bibitem{NLW} S. Kichenassamy,
\emph{Nonlinear Wave Equations}, Dekker, New York, 1995.
\bibitem{FR}
S. Kichenassamy,
\emph{Fuchsian Reduction: Applications to Geometry, Cosmology and Mathematical Physics},
Birkh\"auser, Boston, 2007.
\bibitem{L}
G. Lamb, Elements of Soliton Theory, Wiley, New York, 1980.
\bibitem{rosen}
G. Rosen,
Minimum value for $c$ in the Sobolev inequality
$\norm{\phi^3}\leq c\norm{\nabla\phi}^3$.
SIAM J. Appl. Math. {\bf 21} : 1 (1971) 30--32.
\bibitem{RB} R. Ruffini and S. Bonazzola,
System of self-gravitating particles in General Relativity and the
concept of an equation of state,
Phys. Rev. {\bf 187} : 5 (1969) 1767.
\bibitem{SM} F. E. Schunck and E. W. Mielke, General relativistic
boson stars,
Class. Quantum Grav. {\bf 20} (2003) R301--356.
\bibitem{SS2} E. Seidel and W.-M. Suen,
Dynamical evolution of boson stars: Perturbing the ground state,
Phys. Rev. D {\bf 42} : 2 (1990) 384--403.
\bibitem{SS} E. Seidel and W.-M. Suen,
Oscillating soliton stars,
Phys. Rev. Letters {\bf 66} : 13 (1991) 1659--1662.
\bibitem{Strauss} W. Strauss,
Existence of solitary waves in higher dimension,
Commun. Math. Phys. {\bf 55} (1977) 149--162.
\bibitem{talenti}
G. Talenti,
Best constant in Sobolev inequality
Annali Mat. Pura Appl. {\bf 110} : 1 (1976) 353--372.
\end{thebibliography}
\end{document}